\documentclass[aps,prl,reprint,superscriptaddress,amsmath,amssymb,floatfix,nofootinbib]{revtex4-2}
\usepackage{graphicx,booktabs,bm}
\usepackage[colorlinks=true,linkcolor=blue,citecolor=blue,urlcolor=blue]{hyperref}
\usepackage{orcidlink}
\newcommand{\EE}{\mathbb E}
\newcommand{\T}{\mathcal T}
\newcommand{\M}{\mathcal M}
\newcommand{\subhead}[1]{\par\textit{#1.}\enspace}
\begin{document}
\title{When Can Pulsar Timing Distinguish a Propagation Mass from a Weak Longitudinal Background?}
\author{Kun Zhou~\orcidlink{0009-0001-1099-8623}}
\affiliation{College of Physics, Chongqing University, Chongqing 401331, China}
\affiliation{Department of Physics and Chongqing Key Laboratory for Strongly Coupled Physics, Chongqing University, Chongqing 401331, China}
\author{Jin Li~\orcidlink{0000-0001-8538-3714}}
\email{cqujinli1983@cqu.edu.cn}
\affiliation{College of Physics, Chongqing University, Chongqing 401331, China}
\affiliation{Department of Physics and Chongqing Key Laboratory for Strongly Coupled Physics, Chongqing University, Chongqing 401331, China}

\date{November 5, 2026}
\begin{abstract}
A weak longitudinal background can imitate the angular response of a gravitational-wave propagation mass. We quantify this ambiguity in a compressed Gaussian experiment constructed from the NANOGrav 15-year timing data. At a false-alarm probability of $10^{-3}$, allowing a weak longitudinal component in a massless background costs at least 16.29 percentage points of optimal rejection power for a specified source in a 20-pulsar subarray. Expanding the array to 40 pulsars at the same background level recovers at least 7.62 points. In a separate low-background experiment, we allow the same continuous range of longitudinal amplitudes at zero and positive mass. Across that range and three injected background levels, one test rejects zero mass with at least 97.38\% power at either of two masses separated by 1.08\%. With false assignment limited to $10^{-3}$ at the other mass, the power to identify either mass is at most 1.584\%. A cutoff can thus be identifiable before its position within a frequency cell can be resolved.
\end{abstract}
\maketitle
The angular correlation of pulsar timing residuals depends on the propagation law and polarization of a gravitational-wave background. Dispersion changes both the angular response and the frequencies that can propagate \cite{lee}. Close pulsar pairs are particularly sensitive to longitudinal waves \cite{chamberlin}. In the companion paper we calculate the finite-distance response, including the pulsar terms and the ordered breathing--longitudinal kernels.\footnote{K. Zhou and J. Li, \emph{Pulsar timing correlations from a massive stochastic gravitational-wave background: Exact collinear limits and polarization-dependent distance scaling}, companion manuscript.} A weak longitudinal component can therefore matter on both sides of a test for propagation mass. An increased response alone does not specify which side produced the observed correlations.

Polarization searches in the NANOGrav data \cite{ngpol} and finite-distance calculations within a massive spin-2 model \cite{choi} illustrate two different ways of specifying the competing spectra. Existing PTA mass limits use specified angular correlations or the minimum propagating frequency \cite{wang,wu}. Here we allow independent, incoherent tensor and longitudinal powers in a phenomenological model. We ask how well \emph{any} decision based on a specified set of timing coordinates can reject a family of massless backgrounds. We compare the power of a test that works over the entire family with an upper bound valid for every such test. The comparison also separates evidence against zero mass from the ability to resolve two positive masses.

\subhead{From an angular response to a mass test}
Write $E=m_gc^2$. In the usual positive-mass dispersion law the propagating modes have $hf>E$ and $ck/(2\pi f)=\sqrt{1-(E/hf)^2}$. The overlap reduction function $\Gamma^p_{uv}(f,E)$ describes pulsars $u,v$ before projection onto timing coordinates. For polarization $p$, let $K_p(E)$ denote its spectral covariance after integrating this response over the frequency cells and applying the timing projection on both pulsar indices. Both the shape of $\Gamma$ and the amount of spectral support in a cell can change with $E$.

\begin{figure}[t]
\includegraphics[width=\columnwidth]{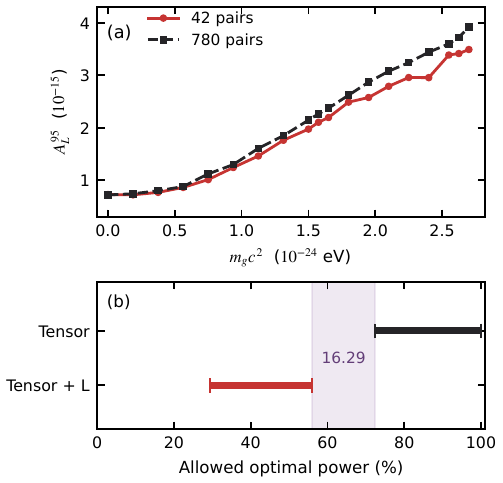}
\caption{Mass-dependent amplitude and loss of discrimination. (a) Near the massless limit, each marker is a measured conditional 95th percentile at $1\,{\rm yr}^{-1}$, with mass fixed and the spectral slope allowed to vary. Lines guide the eye. (b) At a different energy, $E_1$, the intervals bracket the optimal rejection power in a simulated 20-pulsar array at $\rho=1/4$. The purple strip separates a lower bound for the pure tensor null from an upper bound for the mixed null. Its width is a minimum loss, in percentage points.}
\label{fig:loss}
\end{figure}

An empirical example gives the scale of the amplitude ambiguity [Fig.~\ref{fig:loss}(a)]. At 19 fixed masses near the massless limit we infer the conditional 95th percentile of the annual longitudinal strain amplitude under a frozen pair working likelihood. Tensor and longitudinal responses use the same propagation mass. At $E=2.7\times10^{-24}$ eV the percentiles are $3.49\times10^{-15}$ for 42 close pairs and $3.92\times10^{-15}$ for all 780 distinct pairs of the same 40 pulsars. The close-pair percentile is smaller at all 18 positive nodes. A conditional quantile changes with the mass held in the fit. Inverting that curve alone supplies neither the probability of rejecting zero mass nor a calibrated mass interval.

\subhead{The timing experiment}
We use a subset of the NANOGrav 15-year timing release \cite{timing}. The 40-pulsar covariance is constructed in 14 common Fourier cells, with sine and cosine coordinates for each pulsar. A fixed linear map retains six directions per pulsar, chosen using a steep tensor spectrum and the background at a reference mass. The 20-pulsar design takes the corresponding principal block of the same 240-dimensional covariance. The subarray is specified in advance. It is distinct from the 42-edge selection in Fig.~\ref{fig:loss}(a).

The observation is a zero-mean Gaussian vector. With $B$ the fixed shape of the timing-background covariance, its covariance at energy $E$ is
\begin{align}
 C_E={}&\rho B+t^2\{(1-\eta)T_0(E)+\eta T_1(E)\}\nonumber\\
 &+\ell^2 K_L(E).
 \label{eq:model}
\end{align}
$T_0,T_1$ include the reference tensor normalization and have timing-residual spectral indices $13/3$ and $7$, respectively. The longitudinal spectrum has index $13/3$. The weight $\eta$ mixes two power spectra. The amplitudes $t,\ell$ are dimensionless factors in the template normalization. In particular $\ell=0.03$ does not denote a 3\% fraction of the total background power. We compare the massless tensor family $\T$, with $0.8\le t\le1.25$, $0\le\eta\le1$, $\ell=0$, and the family $\M$ with the same ranges and $0\le\ell\le0.03$.

We compare two energies separated by 1.08\% near a frequency-cell edge, $(E_1,E_2)=(2.31146,2.33651)\times10^{-23}$ eV. For measuring loss and array gain we use the reference injection $(t,\ell,\eta)=(1.20,.34,.67)$, with longitudinal amplitude above the massless range. To isolate mass discrimination, our second experiment keeps $(t,\eta)=(1.20,.67)$ and puts the positive source inside the same weak-amplitude range as the massless family.

\subhead{A lower bound on the price of an unknown polarization}
For a positive source $P$ and a family $F$ of massless laws, define the best attainable probability of rejection by
\begin{equation}
 b_F(P)=\sup_{0\le\phi\le1}\{\EE_P\phi:\ \sup_{Q\in F}\EE_Q\phi\le\alpha\}.
 \label{eq:power}
\end{equation}
Here $\alpha=10^{-3}$ for a single field. An admissible test gives a lower bound $L_F$. A single allowed massless distribution $Q_w\in\M$ gives an upper bound $U_w$ on $b_{\M}$ by the Neyman--Pearson lemma \cite{np}: a test valid on the whole family must meet the size constraint at $Q_w$. Since $\T\subset\M$, the loss $\Delta=b_{\T}-b_{\M}$ satisfies
\begin{equation}
 \max(0,L_{\T}-U_w)\le\Delta\le 1-L_{\M}.
 \label{eq:gap}
\end{equation}
To obtain $L_{\M}$ we partition the continuous nuisance parameters into cells. A quadratic score rejects one cell only if its value exceeds a threshold valid throughout the cell. Requiring rejection of every cell controls the false-alarm probability for the union (Appendix). A rule constructed for a given array and $\rho$ is held fixed across the test sources.

At $E_1$ and $\rho=1/4$ the pure-tensor test on the 20-pulsar subarray has power at least 72.32\%. A massless source with $(t,\ell,\eta)=(0.8,0.028373,0)$ limits the power of any size-controlled test on all 120 coordinates to at most 56.03\% [Fig.~\ref{fig:loss}(b)]. The source printed here is rounded; the calculation uses its full precision. The resulting minimum loss of 16.29 points holds within a simultaneous 99\% calibration and sampling guarantee. It persists even if a different test extracts all the information in these coordinates.

We also fixed a single massless competitor, $(t,\ell,\eta)=(.8,.025,0)$, before testing eight new positive sources. A strict power loss could be certified in 423 of 960 combinations of source, array, window, background and energy. In 99 combinations the loss persisted for every competing amplitude in $[.024,.026]$. These counts describe the tested designs.

\begin{figure}[t]
\includegraphics[width=\columnwidth]{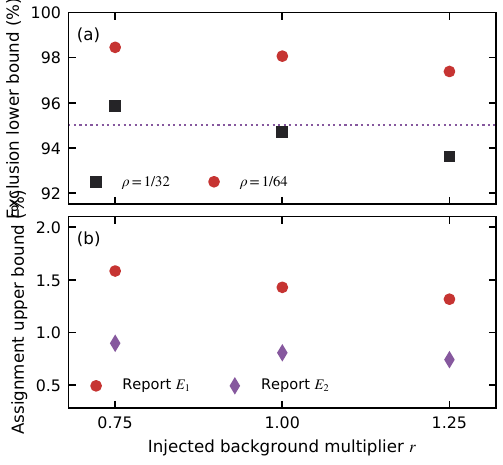}
\caption{Power bounds for a 40-pulsar array with equal longitudinal-amplitude ranges at zero and positive mass. Each marker covers all $0\le\ell\le.03$ at the indicated injected background $r\rho B$. (a) Lower of the two exclusion bounds. Each test controls false alarms for every massless background scale $0.75\le r\le1.25$. (b) Directional assignment bounds at $\rho=1/64$ with size $10^{-3}$ at the competing energy. Only the three marked positive background scales were tested.}
\label{fig:recovery}
\end{figure}

\subhead{What can be recovered?}
At $\rho=1/4$ the reference injection gives a mixed-family rejection power of at least 63.64\% with 40 pulsars, above the optimal upper bound of 56.03\% for its 20-pulsar subarray. The unrounded bounds guarantee a gain of 7.62 percentage points. Keeping the subarray and reducing $\rho$ from $1/4$ to $1/16$ instead guarantees a gain of 7.35 points.

Could exclusion simply reflect the larger longitudinal amplitude in the reference injection? In the 40-pulsar experiment of Fig.~\ref{fig:recovery}, both mass hypotheses allow $0\le\ell\le.03$. The massless background covariance is $r\rho B$ with $r$ free in $[.75,1.25]$. At $\rho=1/64$, a single rule rejects this entire family with at least 97.38\% power at both energies, throughout the weak-amplitude interval and at each of three injected scales $r=.75,1,1.25$. For those same source distributions, a call of $E_1$ with false assignment at $E_2$ limited to $10^{-3}$ has power at most 1.584\%; reversing the energies gives 0.898\%. These bounds allow abstention. At both energies the exclusion test can respond to the loss of propagating power, while identifying the energy requires resolving the smaller change across the cell.

\subhead{Which pairs carry the information?}
An independent injection test compared 42 close pairs with four control graphs of the same size and pulsar degrees. On 65,536 new fields per source, the close pairs outperformed all four controls in nine of nine configurations with longitudinal spectral index $13/3$, and in three of nine with index 2, after correction over 72 paired comparisons. At $E=0.75\times10^{-24}$ eV and $A_L=3\times10^{-15}$, the steep spectrum gave 13,686 calls versus 3,067 for the strongest control. At the same mass and amplitude, the shallow spectrum gave 54 versus 28 and failed the joint criterion. The simultaneous upper false-call bound over 24 new massless sources was 2.811\%; this experiment therefore has a different calibration from Eq.~(\ref{eq:power}).

We propose to rank pair selections by the component of the mass-induced change in their expected cross-correlations orthogonal to changes in nuisance amplitude and slope, in the inverse covariance metric of the pair estimates. The prediction is that this ranking anticipates which pair selections improve discrimination on a different pulsar roster or timing window at equal false-alarm probability. That transfer remains to be tested.

\subhead{Implications}
An array can recover strong evidence against a specified massless family well before it can measure where a cutoff lies within a frequency cell. The scale of that recovery depends on the signal spectrum: at $\rho=1/128$ the attained lower bounds for the alternative injection $(1.02,0.71,0.27)$ at $E_1$ are only 71.9\% (40 pulsars) and 14.2\% (20 pulsars). The background rescalings apply to the complete frozen $B$ and have no assigned conversion to years of observation. The empirical amplitude scan and the simulated decision bounds therefore have different roles. Establishing an observational mass constraint requires calibration with pulsar-specific noise uncertainties and physically justified polarization ranges.

\subhead{Data and analysis} The timing release used here is NANOGrav 15-year v2.1.0, DOI \href{https://doi.org/10.5281/zenodo.16051178}{10.5281/zenodo.16051178}. The reported probabilities refer to the compressed Gaussian experiment defined here. Numerical inputs for the matched comparison and the tabulated figure values are available from the corresponding author. AI-assisted tools were used in drafting, translation, and numerical analysis.
\begin{acknowledgments}
This work was supported by the National Natural Science Foundation of China (Grant Nos. 12575072, 12547101 and 125B2102), the Fundamental Research Funds for the Central Universities Project (Grant No. 2024IAIS-ZD009), and the National Key Research and Development Program of China (Grant No. 2021YFC2203004).

\end{acknowledgments}

\appendix
\section{Construction and checks of the bounds}
\subhead{Frozen coordinates}
For each of the 40 pulsars we diagonalize the reference tensor signal block against the corresponding positive-definite background block in its 28 real Fourier coordinates. We keep the six largest generalized-eigenvalue directions. The reference energy is $2.29904\times10^{-23}$ eV and the reference spectrum has index 7. The two mass hypotheses, including zero where appropriate, use identical coordinate maps within each comparison. The main example uses one frozen noise state (archive label \texttt{p028}) and the window that includes detuning in both the second and third Fourier cells. The normalized tensor kernels in Eq.~(\ref{eq:model}) absorb the square of the common reference amplitude. The subarray inherits its coordinates from the full array. The unit-amplitude longitudinal kernel is specified independently; interpreting a physical theory that ties breathing and longitudinal amplitudes together requires its own covariance.

\subhead{A feasible decision for a continuous null}
Let $X$ be the compressed observation. We whiten at the reference positive source, rank directions by their contributions to the Gaussian relative entropy against a near null selected on a fixed grid, and retain 72 directions. These operations depend only on design covariances. In these coordinates the reference source has unit covariance. We divide $(t^2,\eta,\ell^2)$ into $3\times24\times2$ cells. The null covariance is affine in each of these three coordinates separately, so every interior matrix is a convex combination of the eight corner matrices. In each cell we bracket all corners by positive-definite matrices $C_-\preceq C\preceq C_+$ and use the signed score $X^\top A X$, with $A=C_{\rm rep}^{-1}-I$ for a representative null covariance.

Under $C$, the score has law $\sum_i\lambda_i Z_i^2$, with independent standard normal $Z_i$ and the sorted eigenvalues of $C^{1/2}AC^{1/2}$. An upper envelope uses positive eigenvalues at $C_+$ and negative eigenvalues at $C_-$, preserving their sorted positions. This follows from the min--max principle applied to $v^\top A v/(v^\top C^{-1}v)$ separately on either sign. For the envelope's $1-\alpha$ quantile we take the smaller of a Chernoff upper bound and an upper order statistic from 262,144 Gaussian draws per cell. The latter is simultaneous across 10 designs and 144 cells, allocating failure probability 0.003. Intersecting all cell rejections gives a level-$\alpha$ test for the full union, since a true null lies in at least one cell. The pure-null construction omits the $\ell^2$ coordinate and uses deterministic Chernoff thresholds.

For each of ten designs (two arrays and five known background levels), we freeze the mixed-family rule at the reference source and apply it to six injections (three source spectra and two energies), with 65,536 independent fields per setting. One-sided binomial bounds on the 60 powers have joint failure probability 0.004. The 20-pulsar reference comparison uses a separately simulated pure-null power bound (47,750 rejections in 65,536 positive fields) and the simple competitor $(0.8,0.0283729909108433,0)$. The null and positive likelihood-ratio distributions for that competitor use 524,288 fields each. A lower order statistic for the null threshold and an upper binomial limit on positive exceedances give the conservative upper bound. The pure lower bound and the two steps for the upper bound each use failure probability 0.001. A union bound gives simultaneous coverage of at least 99\% for the main recovery comparisons. This probability concerns numerical calibration and sampling under the model.

\subhead{Resolving neighboring masses}
Let $P,Q$ be centered Gaussian laws on the full retained coordinates, with generalized eigenvalues $\lambda_i$ of $(C_P,C_Q)$. Suppose the probability of reporting $P$ when $Q$ is true is at most $\alpha$. H\"older's inequality applied to $\EE_Q[(p/q)\phi]$ gives, for $s>1$,
\begin{align}
 P(\text{report }P)&\le \alpha^{1-1/s} M_s^{1/s},\nonumber\\
 \log M_s&=-\tfrac12\sum_i\!\left[s\log\lambda_i+
 \log\!\left(1-s(1-\lambda_i^{-1})\right)\right].
 \label{eq:renyi}
\end{align}
We minimize the bound over a grid of admissible orders. A source pair drawn from a larger competing family bounds the assignment power of any test that controls false assignments over the entire family.

\subhead{The same weak-amplitude range at both masses}
For Fig.~\ref{fig:recovery}, the null-cell bounds also cover $r=.75$ and $1.25$ at every corner, hence their full interval. The rule is calibrated once per nominal $\rho$. Order statistics use a total failure budget of .003 over the two designs and their 144 cells each. At each energy, $r=.75,1,1.25$ and $\ell=0,.015,.03$, we generate 65,536 new fields with $(t,\eta)=(1.2,.67)$. Exact one-sided binomial lower limits over these 36 powers share failure budget .007. This gives a separate 99\% simultaneous guarantee for Fig.~\ref{fig:recovery}.

At fixed $E,r,\rho$, write $P_y=N(0,C_0+yK_L)$, $y=\ell^2$, $y_*=.03^2$. With $\lambda_i$ the eigenvalues of $C_0^{-1/2}K_L C_0^{-1/2}$, Pinsker's inequality gives
\begin{equation}
 \sup_{0\le y\le y_*}\|P_y-P_0\|_{\rm TV}\le
 \tau_P=\tfrac12\sqrt{\textstyle\sum_i[y_*\lambda_i-\log(1+y_*\lambda_i)]}.
\end{equation}
Each summand increases with $y$ since its derivative is $y\lambda_i^2/(1+y\lambda_i)$. Subtracting $\tau_P$ from the power lower limit at zero longitudinal amplitude gives panel (a). For the competing energy $Q_y$, a test of size $\alpha$ there has size at most $\alpha+\tau_Q$ at $Q_0$. We evaluate Eq.~(\ref{eq:renyi}) on $(P_0,Q_0)$ at this inflated size and add $\tau_P$ to obtain panel (b). The calculation covers continuous longitudinal amplitude on both sides and three discrete injected background scales.

\end{document}